\documentclass{article}[11pt]
\usepackage{a4wide}
\usepackage{bm}
\usepackage{bbm}
\usepackage{epsfig,graphics}
\usepackage{amsmath,amssymb,amsfonts}
\usepackage{color}
\usepackage{epstopdf}
\usepackage[
colorlinks=true,
linkcolor=black,
breaklinks=true,
urlcolor=blue,
citecolor=green]{hyperref}
\usepackage{cite}

\newcommand{\be}{\begin{equation}}
\newcommand{\ee}{\end{equation}}
\newcommand{\ba}{\begin{eqnarray}}
\newcommand{\ea}{\end{eqnarray}}

\newcommand{\order}[1]{\mathcal{O}\left(#1\right)}

\begin{document}

\title{How the $X(5568)$ challenges our understanding of QCD }

\author{Feng-Kun Guo$^{1,}$\footnote{Email address:
      \texttt{fkguo@itp.ac.cn} },
      Ulf-G. Mei\ss ner$^{1,2,3,}$\footnote{Email address:
      \texttt{meissner@hiskp.uni-bonn.de} }~ and
      Bing-Song Zou$^{1,}$\footnote{Email address:
      \texttt{zoubs@itp.ac.cn} } \\[2mm]
      {\it\small$^1$ Key Laboratory of Theoretical Physics, Institute of
      Theoretical Physics,}\\
      {\it\small  Chinese Academy of Sciences, Beijing 100190, China}\\
      {\it\small$^2$Helmholtz-Institut f\"ur Strahlen- und Kernphysik and Bethe
      Center for Theoretical Physics,}\\
      {\it\small Universit\"at Bonn, D-53115 Bonn, Germany}\\
      {\it\small$^3$Institute for Advanced Simulation, Institut f\"{u}r
       Kernphysik and J\"ulich Center for Hadron Physics,}\\
      {\it\small Forschungszentrum J\"{u}lich, D-52425 J\"{u}lich, Germany}
}
\date{\today}

\maketitle

\begin{abstract}

  We discuss the $X(5568)$ particle recently announced by the D0 Collaboration.
  Several types of models were proposed to explain this structure in the
  literature. As pointed out by Burns and Swanson (arXiv:1603.04366), none of
  them provides a satisfactory description of the observation. We provide
  additional arguments using general properties of QCD, and conclude that the
  observation of the $X(5568)$, if confirmed, poses serious challenges to our
  understanding of nonperturbative QCD.

\end{abstract}

\bigskip

{\it Introduction}.
Very recently, the D0 Collaboration reported the  observation of a narrow peak in
the $B_s^0\pi^\pm$ invariant mass distribution based on  data from $p\bar p$
collisions at $\sqrt{s}=1.96$~TeV~\cite{D0:2016mwd}. They fitted to the data
using an $S$-wave (and a $P$-wave for studying systematic uncertainties)
Breit--Wigner parametrization, and obtained a mass and width of
\begin{equation}
  M_{X(5568)} = \left( 5567.8\pm2.9^{+0.9}_{-1.9}\right)~\text{MeV},\qquad
  \Gamma_{X(5568)} = \left( 21.9\pm6.4^{+5.0}_{-2.5}\right)~\text{MeV}
\end{equation}
with a significance of 5.1~$\sigma$.
This observation has triggered a lot of theoretical calculations and
speculations~\cite{D0:2016mwd,Agaev:2016mjb,Wang:2016tsi,Zanetti:2016wjn,
Wang:2016mee,Chen:2016mqt,
Agaev:2016ijz,Liu:2016xly,Liu:2016ogz,Agaev:2016lkl,Dias:2016dme,Wang:2016wkj,
Agaev:2016urs,He:2016yhd,Jin:2016cpv,Stancu:2016sfd,Burns:2016gvy,Tang:2016pcf},
most of which interpreted the signal as a resonance consisting of two quarks and
two antiquarks, i.e. $\bar b s q\bar q$ ($q=u,d$), either of a compact
tetraquark or meson-meson molecular structure. There is also an explanation
using the so-called triangle singularity~\cite{Liu:2016xly} by observing that the peak is
located only about 13~MeV above the $B_s^*\pi^\pm$ threshold, where the
$B_s^*\pi\to B_s\pi$ rescattering is required. A warning message was delivered
in Ref.~\cite{Burns:2016gvy}, where the difficulties of interpreting the signal
as a tetraquark, a hadronic molecule or a threshold effect were discussed. In
this short note, we want to sharpen the conclusions of Ref.~\cite{Burns:2016gvy}
by using very general arguments from quantum chromodynamics (QCD) which include
chiral symmetry and heavy quark symmetry. We will argue that if the $X(5568)$
will be confirmed with the reported properties, then it would pose
serious challenges to our understanding of nonperturbative QCD.

{\it Chiral symmetry}.
The QCD Lagrangian has a SU$(N_f)_L\times$SU$(N_f)_R$
symmetry, where $N_f=2$ or 3 is the number of light quark flavors, which is
spontaneously broken to the vector subgroup SU$(N_f)_V$. Although the
spontaneous breaking of chiral symmetry has not been proven theoretically,
strictly speaking, there are strong evidences for it from both experiment and
lattice QCD simulations. As a result, of this spontaneous symmetry breaking,
there are $N_f^2-1$ Goldstone bosons which are identified as the
lightest pseudoscalar multiplet. This explains why the pion mass,
$M_\pi\simeq138$~MeV, is much lighter than the mass of any other hadron. Chiral
perturbation theory (CHPT)~\cite{Weinberg:1978kz,Gasser:1983yg} is the
low-energy effective field theory for QCD based on chiral symmetry and its
spontaneous breaking. At leading order, one has the Gell-Mann--Oakes--Renner
relation
$M_\pi^2 = 2 B m_q\propto m_q$,  with $m_q$ the light quark mass and
$B$ a positive constant. It is clear that
the pion mass vanishes in the chiral limit $m_q\to 0$.
In CHPT, one counts $M_\pi=\order{p}$ with $p\ll M_R$ being a small momentum
scale where $M_R$ denotes the mass of the lowest resonance, which can not be
described in a perturbative expansion like CHPT.
Any hadron other than the Goldstone bosons has a nonvanishing mass in the chiral
limit, and this mass is of order $\order{p^0}$. This can be generalized to argue that
introducing any additional  quark-antiquark ($q\bar q$) pair   will add a mass
of  $\delta m_{q\bar q}=\order{M_R}\gg M_\pi$. The $B_s$ is the ground state of
$\bar b s$ systems, and any excitation will increase the mass. We thus expect
that a $\bar b s \bar q q$ tetraquark to have a mass
\begin{equation}
M_{\bar b s\bar q q} \gtrsim M_{B_s} + M_R .
\end{equation}
If we estimate $M_R$ by the $f_0(500)$ meson mass, which is the lowest meson
resonance, then we get $\sim5.9$~GeV. It is much above the observed mass of the
$X(5568)$, consistent with the expectation in Ref.~\cite{Burns:2016gvy} from the
point of view of constituent quark model.
The only possibility evading such a large mass is to generate a state from the
interaction between bottom mesons and Goldstone bosons, more precisely, pions.
If the interaction between matter field and pseudo-Goldstone boson is strong,
then it is possible to generate a state close to the threshold. Since the
reported mass is not far from the $B_s\pi$ threshold, we discuss only the
$S$-wave case. However, the low-energy pionic interaction is weak because of
chiral symmetry. Furthermore, the leading order interactions in the chiral
expansion in both of the $B_s\pi$ and isovector $B\bar K$ channels vanish, see,
e.g., Ref.~\cite{Guo:2009ct}.
It is possible to generate a pole from coupled-channel effects. But then the
pole should not be located only 60~MeV above the $B_s\pi$ threshold.
In fact, in the charmed sector, a pole can be generated by the isovector
coupled-channel $D_s\pi$--$DK$ dynamics~~\cite{Guo:2009ct}, but it is deep
in the complex energy plane with a real part about 200~MeV higher than the
$D_s\pi$ threshold.

{\it Heavy quark symmetry}. In the heavy quark limit, for any hadron containing
a single heavy quark, the heavy quark mass does not play a role in the dynamics
(see, e.g., Ref.~\cite{Manohar:2000dt}). This leads to heavy quark flavor
symmetry (HQFS). If the $X(5568)$ is a hadron resonance, either being a compact
tetraquark or of meson-meson type, it should have a charmed partner.
The HQFS breaking effects are proportional to
$(1/m_c-1/m_b)$~\cite{Manohar:2000dt}. We therefore expect the charmed partner
of the $X(5568)$ to have a mass around
\begin{equation}
M_{X_c} = M_{X(5568)} - \bar M_{B_s} + \bar M_{D_s} +
\order{\Lambda_\text{QCD}^2\left(\frac1{m_c}-\frac1{m_b}\right)}\simeq
\left(2.24\pm0.15\right)~\text{GeV},
\end{equation}
where $\bar M_{B_s}=5.403$~GeV and $\bar M_{D_s}=2.076$~GeV are the
spin-averaged masses of the pseudoscalar and vector heavy-strange mesons. Within
this mass range, the only candidate is the $D_{s0}^*(2317)$ which was discovered
by the BaBar Collaboration in the $D_s\pi$ invariant mass
distribution~\cite{Aubert:2003fg} and has a width of less than
3.8~MeV~\cite{Agashe:2014kda}.
It is generally believed that the $D_{s0}^*(2317)$ is an isospin scalar (isoscalar) meson,
and its decay into the $D_s$ and pion is due to isospin breaking effects, which
has been proposed~\cite{Faessler:2007gv,Lutz:2007sk,Cleven:2014oka}
to be used to discriminate the hadronic molecular
interpretation~\cite{Barnes:2003dj,
Kolomeitsev:2003ac,Hofmann:2003je,Guo:2006fu,Zhang:2006ix,Gamermann:2006nm} from
other possibilities. The width of the $X(5568)$ is about 20~MeV. The only
allowed possible strong decay modes are the $B_s\pi$ and $B_s^*\pi$, both of
which are isovector. Therefore, were the $X(5568)$ a hadronic resonance, its
isospin should be 1, and hence the $D_{s0}^*(2317)$ cannot be its charmed
partner.
One needs to answer the question why the charmed partner of the $X(5568)$, with
a width ideal for observation, has not been observed in the same processes where the very
narrow $D_{s0}^*(2317)$ was observed.

{\it Could the $X(5568)$ be due to $B_s^*\pi$ threshold effects?} Liu and Li
suggested to explain the peak as a triangle singularity effect observing that it
is close to the $B_s^*\pi$ threshold~\cite{Liu:2016xly}. Their model
involves the rescattering from $B_s^*\pi$ to $B_s\pi$. This mechanism was questioned in
Ref.~\cite{Burns:2016gvy} as such a process is too weak as it is in a
$P$-wave and no flavor exchange is possible. Here we will show to what extent
such a rescattering is suppressed. Around the $X(5568)$ peak, the pion momenta
in both $B_s\pi$ and $B_s^*\pi$ are low.  Thus, we can describe such a process
by an effective chiral Lagrangian respecting heavy quark symmetry. We denote the
$B_s$ and $B_s^*$ spin multiplet by a superfield $H_s=\vec B_s^*\cdot
\vec\sigma + B_s$~\cite{Hu:2005gf}, where $\vec\sigma$ are Pauli matrices,
which transforms under parity ($\mathcal{P}$) and heavy quark spin
($\mathcal{S}$) transformations as
\begin{equation}
  H_s \overset{\mathcal{P}}{\to} -H_s, \qquad H_s \overset{\mathcal{S}}{\to} S
  H_s,
\end{equation}
where $S$ is the rotation matrix acting on the heavy quark spin, and does not
commute with the Pauli matrices. The lowest order operators, preserving heavy
quark spin symmetry, in the chiral expansion that have a nonvanishing
contribution to $B_s^*\pi^\pm\to B_s\pi^\pm$ are
\begin{equation}
i\,\epsilon_{ijk}\left\langle H_s^\dag \partial^i H_s \sigma^j
\right\rangle \partial^k \left( \chi_+ \right)_{aa}, \qquad
i\,\epsilon_{ijk}
\left\langle H_s^\dag H_s \sigma^i \right\rangle
\left( \left[u^j, u^k\right] \chi_+ \right)_{aa} ~,
\end{equation}
where $\langle,\rangle$ denotes the trace in the spinor space and $a$ is the
light flavor index. Here $\chi_+$ and $u^i$ are the usual building blocks of
chiral Lagrangians. They contain an even number and an odd number of pion
fields, respectively, and read
$\chi_+ 
= 2 \chi -\{\pi,\{\pi,\chi\}\}/(4F^2)+\ldots$ and
$u^i 
=-\partial^i \pi/F+\ldots$,
where
$\chi=2B\cdot\text{diag}(m_u,m_d)$, $F$ is the pion decay constant in the
chiral limit, and $\pi$ is the usual $2\times2$ matrix for the pion isospin
triplet. These operators are doubly suppressed:
(1) the lowest order
possible operators in the chiral expansion, as given above, are of
$\order{p^4}$, the next-to-next-to-next-to-leading order, in the chiral
expansion for the interaction between Goldstone bosons and matter fields, and
thus highly
suppressed; (2) the bottom-strange mesons and pions do not have a common flavor,
and the interaction is thus OZI suppressed or $1/N_c^2$ suppressed (a
relative $1/N_c$ suppression in comparison with those scattering processes
with a common quark flavor), with $N_c$ the number of colors.
These strong suppressions make less likely explaining the
observed peak by invoking $B_s^*\pi\to B_s\pi$ rescattering.

{\it Summary}.
To summarize, our arguments based on chiral symmetry and heavy quark symmetry
support the analysis in Ref.~\cite{Burns:2016gvy} that it is hard to explain the properties
of the $X(5568)$ using either tetraquark, hadronic molecule, or
threshold-effect models. If the observation of the $X(5568)$ is confirmed, it
would have an important impact on the understanding of nonperturbative QCD. We
thus suggest to search for it in other processes such as the dipion decays of
excited bottom-strange mesons, e.g.,
$B_{s2}^*(5840)\to B_s\pi\pi$, and search for its charmed partner using the huge
data sets of $B$ factories. In fact, the LHCb Collaboration did not see a
signal corresponding to the $X(5568)$ based on their $pp$ collision
data~\cite{LHCb}.

\medskip

\section*{Acknowledgments}

We would like to thank Xiao-Hai Liu and Eulogio Oset for useful comments, and
Marek Karliner for informing us about the LHCb conference note. This work is
supported in part by DFG and NSFC through funds provided to the Sino-German CRC
110 ``Symmetries and the Emergence of Structure in QCD'' (NSFC Grant No.
11261130311), by the Thousand Talents Plan for Young Professionals, and by the
Chinese Academy of Sciences President's International Fellowship Initiative
(Grant No. 2015VMA076).

\end{document}